\documentstyle[twocolumn]{jpsj}

\def\runtitle{
Alternating Spin Chain with $S=1/2,~1$ in Magnetic Fields
}
\def\runauthor{M. Fujii, S. Fujimoto and N. Kawakami}

\title{
Magnetic and Critical Properties of \\
Alternating Spin Chain with $S=1/2,~1$ in Magnetic Fields
}

\author{Mitsuru Fujii$^{1}$, Satoshi Fujimoto$^{2}$ 
and Norio Kawakami$^{1,3}$
}

\inst{
$^{1}$ Department of Applied Physics, 
Osaka University, Suita, Osaka 565, Japan  \\
$^{2}$ Department of Physics,
Kyoto University, Kyoto 606, Japan \\
$^{3}$ Department of Material and Life Science, 
Osaka University, Suita, Osaka 565, Japan
}

\recdate
{\today}

\abst{
We study an integrable spin chain with 
an alternating array of spins $S=1/2,~1$ 
in external magnetic fields using the Bethe ansatz exact solution. 
The calculated magnetization possesses a cusp structure 
at a critical magnetic field $H=H_{C}$,
at which the specific heat shows a divergence property.
We also calculate finite-size corrections to the 
energy spectrum, and obtain the 
critical exponents of correlation functions
with the use of conformal field theory (CFT). 
Low-energy properties of the model are described 
by two $c=1~U(1)$ CFTs in $H<H_{C}$ and one $c=1~U(1)$ CFT 
in $H>H_{C}$. 
}
\kword
{alternating spin chain, exact solution, conformal field theory}

\begin{document}
\sloppy
\maketitle
It is now well established that the antiferromagnetic
quantum spin chains  with  half-integer spins 
have massless excitations, whereas 
those with integer spins have massive ones (Haldane gap),
although the ground state of both cases are 
known to be singlet.
Experimentally, another interesting class of 
quantum spin systems have been found, i.e. 
the spin systems with an alternating array of 
integer and half-integer spins
both for ferromagnetic and antiferromagnetic cases\cite{exp}.
The systems found so far have the magnetically ordered
ground state even for the antiferromagnetic case,
for which quantum fluctuations may not play a crucial role. 
It is interesting to explore what happens 
if  the magnetic order disappears, 
for example, in the presence of certain interactions 
beyond nearest neighbor ones. If such a system is
realized, quantum fluctuations are expected to become
important. This problem may thus provide us with 
a new interesting paradigm of quantum spin chains.  

As a first step to address this problem of
alternating spin chains with a 
singlet ground state, it may be 
interesting to study an integrable spin model 
\cite{d-1,d-2,m-1,m-2} whose ground state is singlet. 
In this paper, we study the integrable alternating spin model of 
$S=1/2, 1$, with  a special attention to its
critical properties in magnetic fields. We 
systematically evaluate the magnetization, the specific 
heat and the critical exponents of correlation functions,
and discuss their characteristic properties.

We consider a system of $N$ spins 
$\frac{1}{2} \mbox{\boldmath $\sigma$}_{2}$,
$\frac{1}{2} \mbox{\boldmath $\sigma$}_{4}$,
$\cdots$,$\frac{1}{2} \mbox{\boldmath $\sigma$}_{2N}$ 
of spin-$\frac{1}{2}$ and $N$ spins 
$\mbox{\boldmath $S$}_{1}$,
$\mbox{\boldmath $S$}_{3}$,
$\cdots,\mbox{\boldmath $S$}_{2N - 1}$ 
of spin-$1$ in an external magnetic field $H$, 
the Hamiltonian for which is given by\cite{d-1}
\begin{eqnarray}
{\cal H} & = &
 \frac{1}{9} \sum_{n=1}^{N} 
\left\{ \left(2
\mbox{\boldmath $\sigma$}_{2n} \cdot \mbox{\boldmath $S$}_{2n+1} 
+ 1 \right)
\left(2
\mbox{\boldmath $\sigma$}_{2n+2} \cdot \mbox{\boldmath $S$}_{2n+1} 
+ 3 \right) \right.
\nonumber \\
& & \left.
{} + 
\left(2
\mbox{\boldmath $\sigma$}_{2n} \cdot \mbox{\boldmath $S$}_{2n-1} 
+ 1 \right)
\left[\left(
\mbox{\boldmath $\sigma$}_{2n-1} \cdot \mbox{\boldmath $\sigma$}_{2n+1} 
+ 1 \right) 
\left(2
\mbox{\boldmath $\sigma$}_{2n} \cdot \mbox{\boldmath $S$}_{2n+1} 
+ 1 \right) 
+ 2 \right]\right\} 
- H S^{z} ~.
\label{hamil}
\end{eqnarray}
where   periodic boundary conditions are imposed: 
$\mbox{\boldmath $\sigma$}_{2n} = 
\mbox{\boldmath $\sigma$}_{2n+2N} $
and
$\mbox{\boldmath $S$}_{2n-1} =
\mbox{\boldmath $S$}_{2n-1+2N} $.
Note that the Hamiltonian contains rather complicated 
next-nearest-neighbor interactions in addition to
nearest-neighbor Heisenberg-type interactions, in order for 
the model to be integrable. 
The ground state of the model is known to be 
singlet \cite{d-1,d-2,m-1,m-2}, for which,
as will be shown below, 
the antiferromagnetic correlation is dominant.
In spite of its ugly appearance (\ref{hamil}),
we wish to note that this system may provide a starting model 
to discuss fundamental properties of alternating spin
chains with quantum disordered (singlet) ground states.
In the end of  the paper, we shall give some discussions 
about what would be expected beyond this special model.

Let us first summarize the basic equations 
necessary for the following analysis.
The Bethe equation obtained via the 
quantum inverse scattering method for (\ref{hamil})
is \cite{d-1,d-2,m-1,m-2}
\begin{eqnarray}
\left(
\frac{\lambda_{j} - \frac{i}{2}}{\lambda_{j} + \frac{i}{2}}
\right)^{\frac{L}{2}}
\left(
\frac{\lambda_{j} - i}{\lambda_{j} + i}
\right)^{\frac{L}{2}}
 = 
- \prod^{M}_{l=1} 
\frac{\lambda_{j} - \lambda_{l} - i}{\lambda_{j} - \lambda_{l} + i} ~,
\label{beth}
\end{eqnarray}
where $L = 2N$ is assumed to be a multiple of 4.
Here we have taken the fully polarized state as a reference 
state, and denotes the number of 
``particles" by $M$, which are created by 
the spin-lowering operator. So, the $z$-component of the total 
spin is to be  $S^z = 3N/2 - M$.
One can see two kinds of phase factors 
reflecting the mixture of spin 1 and 1/2
in the left-hand side of (\ref{beth}).
The total energy is
\begin{eqnarray}
E & = &
\frac{7}{6}N - \sum_{j=1}^{M} 
\left(
\frac{1}{\lambda_{j}^{2} + \frac{1}{4}} +
\frac{2}{\lambda_{j}^{2} + 1}
\right)
- H S^{z} ~
\end{eqnarray}
and the total momentum is given by
$P = {\rm Im} \sum_{j=1}^{M}
\log [(\lambda_{j} - \frac{i}{2})(\lambda_{j} - i)] $.
So far, thermodynamic properties 
of the model have been investigated, revealing 
some properties distinct from those of the ordinary 
Heisenberg spin chain \cite{d-1,d-2,m-1,m-2}. Also, 
conformal properties have been discussed to some extent: 
the system at zero field is classified as 
CFT of central charge $c=2$ \cite{d-2,m-1}, and is
conjectured to be described by 
level-1 SU(2) Wess-Zumino-Witten models\cite{m-1}.
In what follows, we systematically 
study the model in the presence of magnetic fields, and 
show that below a certain critical field $H_C$,
low-energy properties of the model are described 
by two $c=1~U(1)$ CFTs, while 
for $H>H_{C}$ by one $c=1~U(1)$ CFT. 
Accordingly, the system shows quite different 
critical behaviors 
for low fields $H<H_{C}$ and high fields $H>H_{C}$.

We start with the calculation of the 
magnetization curve. To this end we 
need the Bethe equations in the thermodynamic limit.
For large enough $L$, the ground state properties 
can be described by $M_{\rm R}$
 real rapidities  $\lambda^{\rm R}_{j}$
and 2-string complex rapidities $\lambda^{\alpha}_{j}$,
\begin{eqnarray}
\lambda^{\alpha}_{j}
& = & \lambda^{\rm C}_{j} + \frac{1}{2} i 
\left(3 - 2 \alpha\right) ~~(\alpha = 1,2) ~,
\label{lam}
\end{eqnarray}
where $\lambda_{j}^{\rm C}$ 
($j=1,2, \cdots, M_{\rm C}$)
is a real number which labels the center 
of the $2$-string \cite{d-1,d-2,m-1,m-2}.
 The Bethe equation (\ref{beth})
is thus converted into integral equations for the density functions 
$\rho^{\rm R}(\lambda^{\rm R})$ and
$\rho^{\rm C}(\lambda^{\rm C})$
for these rapidities
\begin{eqnarray}
 \left(\begin{array}{c}
\rho^{\rm R} \\
\rho^{\rm C} 
\end{array} \right)
& = & 
\left( \begin{array}{c}
\rho^{\rm R}_{0} \\
\rho^{\rm C}_{0}    
\end{array} \right)   \nonumber \\ 
&- & 
\left( \begin{array}{cc}
K_{\rm RR} & K_{\rm CR} \\
K_{\rm RC} & K_{\rm CC}  
\end{array} \right)
\otimes
\left(\begin{array}{c}
\rho^{\rm R}\\
\rho^{\rm C}\end{array}\right) ~,
\label{vec}
\end{eqnarray}
where the integral kernels are
given by $K_{\rm RR}(\lambda) = K_{1}(\lambda)$,
$K_{\rm CC}(\lambda) =
\left\{2 K_{1}(\lambda) +K_{2}(\lambda)\right\}$, 
$K_{\rm CR}(\lambda) =K_{\rm RC}(\lambda) =
\left\{K_{1/2}(\lambda) +K_{3/2} 
(\lambda)\right\}$, respectively, in terms of the function
$K_{n}(x) = (n/\pi)[n^{2} + x^{2}]^{-1}$. 
The driving terms read
\begin{eqnarray}
\left(
\begin{array}{c}
\rho^{\rm R}_{0}
\\
\rho^{\rm C}_{0}
\end{array}
\right) = \frac{1}{2}
\left( 
\begin{array}{c}
K_{1}+ K_{1/2}\\
K_{1/2} + K_{1}+ K_{3/2}
\end{array}
\right),
\label{drv1}
\end{eqnarray}
and the symbol $\otimes$ denotes the convolution 
integral in the finite 
range of $[\Lambda_\alpha^{-},\Lambda_\alpha^+]$
respectively for the sectors of  $\alpha$=R, C.

By calculating  the above integral
equations (\ref{vec}) numerically\cite{numeric}, we have computed 
the magnetization ${\cal M}$$(\equiv S^z)$
as a function of the magnetic field.
The results are shown in Fig.\ref{magn}. 
We can see that the magnetization begins
linearly in magnetic fields (finite spin susceptibility)
since the ground state is singlet, and increases smoothly up to 
the critical field $H_C$, where we encounter
a characteristic cusp structure \cite{com}.
The existence of the cusp at $H=H_{C}$ indicates that 
the magnetization for $H<H_{C}$ consists of two contributions,
i.e. ${\cal M}={\cal M}_{\rm R}+{\cal M}_{\rm C}$,
and one of them, ${\cal M}_{\rm C}$, saturates at $H=H_{C}$.
For $H>H_C$, the remaining massless spin excitations 
still contribute to the magnetization, and then at $H=6$
the system is completely polarized.
These characteristic properties directly reflect  that 
the system consists of two kinds of different spins.
The anomaly at $H =H_C$ also appears in the specific heat.
We show the computed results for the coefficient $\gamma$
of the $T$-linear specific heat in Fig.\ref{gamma}.
The divergence property for $H \rightarrow H_C$ 
implies that the velocity for one of spin
excitations  tends to zero when 
the system approaches the critical field $H_C$,
since the specific heat coefficient $\gamma$ is 
inversely proportional to the velocity. 
To see this point clearly, we have plotted  
two velocities $v_{\rm R}$ and $v_{\rm C}$ in Fig.\ref{velo}.
It is  seen that $v_{\rm C}$ indeed vanishes at $H=H_{C}$. 
Therefore, we can say that
there exist two species of spin excitations which behave
differently in  magnetic fields.
They have the distinct pseudo-Fermi surfaces
$p^{\rm R}_{\rm F}$ and $p^{\rm C}_{\rm F}$, both of them 
decrease with the increase of 
magnetic fields. In particular, at $H=H_C$,
$p^{\rm C}_{\rm F}$ vanishes and the corresponding spin
excitation becomes massive. In what follows we refer 
to these spin excitations as
spinons in the R- and C-sectors respectively.

In order to further study critical properties of
the model in magnetic fields, we now calculate the finite-size 
spectrum. Following a standard technique\cite{woy} we obtain
the corrections to the energy
\begin{eqnarray}
E & = & L \varepsilon_{\infty}
(\pm \lambda_{\rm R}^{0},\pm \lambda_{\rm C}^{0})  
- \frac{\pi v_{\rm R}}{6 L}
- \frac{\pi v_{\rm C}}{6 L} \nonumber \\
& +& \frac{2 \pi v_{\rm R}}{L}(
\Delta_{\rm R}^{+} +\Delta_{\rm R}^{-})
 + \frac{2 \pi v_{\rm C}}{L}
(\Delta_{\rm C}^{+} +\Delta_{\rm C}^{-}) ~.
\label{corr}
\end{eqnarray}
where $\varepsilon_{\infty}$ is the bulk energy density
in the thermodynamic limit.
Also, the finite-size corrections to the momentum are 
\begin{eqnarray}
P & = &
\sum_{\alpha= {\rm R, C}}[- 2  p^{\alpha}_{\rm F}
\Delta D_{\alpha} - \frac{2 \pi}{L}
(\Delta_{\rm \alpha}^{+}- 
\Delta_{\rm \alpha}^{-})]
\label{mom}
\end{eqnarray}
where ``the Fermi momenta'' for two spinons
are given by $p^{\rm R}_{\rm F}=\pi/4-\pi{\cal M}_{\rm R}/L$ 
and $p^{\rm C}_{\rm F}=\pi/2-\pi{\cal M}_{\rm C}/L$
with ${\cal M}_{\rm R}=L/4-M_{\rm R}$ 
and ${\cal M}_{\rm C}=L/2-2M_{\rm C}$. Here
$\Delta D_{\alpha}$ labels an excitation across the
two  Fermi points, which carries the large momentum transfer 
$2 p^{\alpha}_{\rm F} \Delta D_\alpha$.

We note that both of the above expressions (\ref{corr})
and (\ref{mom}) completely fit in with the finite-size scaling 
formulae in  CFT\cite{fss}. Namely, the corrections 
$-\pi v_{\alpha}/(6L)$ to
the ground state energy in (\ref{corr}) 
indicate that two kinds of spinons are described by 
two $c=1$ CFTs. Also, the conformal dimensions 
$\Delta_{\alpha}^{\pm}$ which 
enter  the excitation energy  as
well as the momentum are of desirable forms 
predicted by CFT, 
\begin{eqnarray}
\Delta_{\alpha}^{\pm} & = &
\frac{1}{2} \left\{ \left(
\frac{\xi_{\beta\beta} \Delta M_{\alpha} -
\xi_{\alpha\beta} \Delta M_{\beta}}
{2 {\rm det} [\xi]} \right) 
 \pm  \left(\xi_{\rm {\alpha\alpha}} \Delta D_{\rm \alpha} +
\xi_{\beta\alpha} \Delta D_{\beta} \right) \right\}^{2} 
+ n^{\pm}_{ \alpha}, 
\label{dimension}
\end{eqnarray}
for $\alpha={\rm R, C}$.
Here the so-called dressed 
charge matrix  $\xi_{\alpha\beta}$ \cite{dressed}
is determined by the following integral equations,
\begin{eqnarray}
\xi & = &
I - K \otimes \xi ~,
\end{eqnarray}
where $K$ is the $2\times 2$ matrix of integral kernels 
which has appeared in eq.(\ref{vec}), and $I$ is the 
identity matrix.  The above dressed charge matrix 
determines the critical exponents 
of the present model in magnetic fields.
Note that the conformal dimensions (\ref{dimension}) 
both for R- and C-sectors have a typical form
inherent in $c=1$ Gaussian CFT with U(1) symmetry, which 
has previously appeared in various  contexts, for instance,
in the analysis of interacting electron systems
\cite{electron}.
Recall here that for $H>H_C$, spinons of 
the C-sector become massive 
with vanishing $v_{\rm C}$. Therefore we can see
that low-energy properties of the present 
model in magnetic fields  are described by two $c=1$ CFTs 
for  $0<H<H_{C}$ and one $c=1$  CFT for
$H \geq H_{C}$.

Having specified the critical behavior 
in magnetic fields by CFT, we now  
read  critical exponents of correlation 
functions from the conformal dimensions (\ref{dimension})
by suitably choosing the quantum numbers.
For this purpose, we should note 
that the quantum numbers in (\ref{dimension}) are subject to 
the selection rules
\begin{eqnarray}
\Delta D_{\rm R} =
\frac{\Delta M_{\rm R}}{2},    
~~~\Delta D_{\rm C} & = &
\frac{\Delta M_{\rm C}}{2} \pmod{1} ~,
\end{eqnarray}
which directly reflect that spinons can be 
regarded as interacting bosons.

For $H<H_{C}$ both of two kinds of spinons
control the long-distance behavior of
correlation functions, since 
$v_{\rm R}$ and $v_{\rm C}$ have finite values. 
Thus choosing  the quantum numbers as
\begin{eqnarray}
(\Delta M_{\rm R},\Delta M_{\rm C},
\Delta D_{\rm R},\Delta D_{\rm C}) 
&  = & \left\{
\begin{array}{c}
(0,0,1,0) \\
(0,0,0,1)
\end{array}
\right. ,
\label{choice}
\end{eqnarray}
we find  the spin-spin correlation function to
have the asymptotic form,
\begin{eqnarray}
\displaystyle
\chi_{z}(x) & \sim &
\frac{a_{0}}{x^{2}} + 
\frac{a_{1} \cos (2 p_{\rm F}^{\rm R} x)}
{ x^{ \alpha_s^{\rm R}}} +
\frac{a_{2} \cos (2 p_{\rm F}^{\rm C} x) }{x^{ \alpha_s^{\rm C}}}
\label{spinspin}
\end{eqnarray}
Note that 
the spin-spin correlation function has two oscillating terms,
reflecting the presence of two-types of spinons. 
By inserting (\ref{choice}) into (\ref{dimension}), and 
using the formula 
$\alpha=2\sum_{p=\pm}[\Delta_{\rm C}^{p}+
\Delta_{\rm R}^{p}]$,
we obtain critical exponents for the oscillating parts,
\begin{eqnarray}
\alpha_{s}^{\rm R}
= 2 (\xi_{\rm {RR}}^{2} + \xi_{\rm {RC}}^{2}) ~, ~~~~~
\alpha_{s}^{\rm C}
= 2 (\xi_{\rm {CR}}^{2} + \xi_{\rm {CC}}^{2}) ~.
\end{eqnarray}
The numerical results for $\alpha_{s}^{\rm R}$ 
and $\alpha_{s}^{\rm C}$  are
plotted as a function of the magnetic field  in Fig.\ref{exp}. 
In weak magnetic fields, $\alpha_s^{\rm C}$ represents how
dominant the antiferromagnetic fluctuations are, because
$p_{\rm F}^{\rm C} \sim \pi/2$.  
Since the value of $\alpha^{\rm C}_s$
is close to unity in weak fields, 
we can see that the antiferromagnetic 
correlation is indeed dominant there,
 but is gradually suppressed with
the increase of the magnetic filed.
For $H\geq H_C$, spinons of the C-sector become massive,
and remaining massless spinons in the R-sector solely determine 
the critical behavior of the system.
Accordingly, the third term of the right-hand 
side in eq.(\ref{spinspin}) vanishes. Also,
 the critical exponent $\alpha_s^{\rm R}$ exhibits 
a discontinuity at $H=H_C$. This decrease in the 
critical exponent $\alpha_s^{\rm R}$ may imply that 
the corresponding spin correlation may be enhanced 
to some extent in magnetic fields.  Similar phenomena have 
been indeed observed in a different context where spin correlations
are enhanced by decreasing the multiplicity of 
bands \cite{tj}.  As seen from Fig.\ref{exp}, however, 
the corresponding exponent is larger than unity, 
so this fluctuation may not become dominant.
We note here that a logarithmic correction 
should enter the correlation functions at $H=0$ since the symmetry of 
the system is enhanced from $U(1)$ to  $SU(2)$ 
at $H=0$, although it does not appear in the present case
with finite magnetic fields. 

Here, following the arguments of Sachdev\cite{sac}, 
we would like to briefly discuss the temperature dependence
of nuclear magnetic relaxation rates which may be
an experimental probe to study 
low-energy properties of the system.
Since the low-temperature behavior of the model is mainly determined by
spinons mentioned above, we can obtain the leading temperature
dependence of the nuclear spin-lattice relaxation rates $1/T_1$
and the nuclear spin-spin relaxation rates $1/T_{2G}$ from
eq.(\ref{spinspin})\cite{sac}, 
\begin{eqnarray}
\frac{1}{T_1}&\sim& \frac{C_1}{T^{1-\alpha_s^{\rm R}}}
+\frac{C_2}{T^{1-\alpha_s^{\rm C}}}+C_3 T ~, 
\label{t1} \\
\biggl(\frac{1}{T_{2G}}\biggr)^2&\sim& \frac{D_1}{T^{3-2\alpha_s^{\rm R}}}
+\frac{D_2}{T^{3-2\alpha_s^{\rm C}}} ~.\label{t2g}
\end{eqnarray}
Combining these expressions with 
the numerical results for the critical exponents shown in 
Fig.\ref{exp}, we see that $1/T_1$ behaves like 
$\sim 1/T^{1-\alpha_s^{\rm C}}$ at low temperatures for $H<H_{C}$.
However for $H\geq H_C$ this term vanishes and 
the low temperature behavior of $1/T_1$ 
is determined by the term $\sim 1/T^{1-\alpha_s^{\rm R}}$.
The temperature dependence of $1/T_{2G}$ for $H<H_{C}$ is also
governed by spinons in the C-sector and given by 
$1/T^{3/2-\alpha_s^{\rm C}}$.
For $H\geq H_C$ $1/T_{2G}$ behaves 
like $\sim 1/T^{3/2-\alpha_s^{\rm R}}$.

We have been concerned with the rather special model with 
alternating spins in this paper. Before  concluding the paper,
we wish to briefly mention what would be expected
if the system were away from the solvable point with  
coupling constants being changed. Unfortunately there have not been 
systematic attempts to this problem so far, so that
we summarize here two possible disordered ground 
states predicted by the present analysis. (i) A possibility is that   
one of the $c=1$ fields becomes massive in the presence of
some relevant perturbations.
In this case, we can see from the present results that
low-field properties are classified 
by $c=1$ CFT, and then the magnetization begins linearly.
What is  characteristic for this case is that a cusp structure
in Fig.1  may not be observed.
(ii) Another interesting possibility may be that 
both of two spinons become massive, and hence
 the system shows properties like a Haldane-gap system,
for which there is no magnetization in low fields. 
These states, as well as the present massless state,
should compete with the magnetically ordered state.
In order to fully understand these problems, 
an extensive study with the aid
of bosonization methods and numerical methods
would be quite helpful. This issue is now under consideration.

In summary we have investigated magnetic and critical 
properties of an integrable 
alternating spin chain in external magnetic fields.
The critical properties in magnetic fields are
classified in terms of CFT.
Although the present analysis is based on a rather special 
model, it would provide a plausible
guideline to systematically explore characteristic properties of   
this new class of quantum spin chains.

%


\begin{figure}
\caption{Magnetization as a function of the magnetic field.}
\label{magn}
\end{figure}
\begin{figure}
\caption{Specific-heat coefficient as a 
function of the magnetic field.}
\label{gamma}
\end{figure}
\begin{figure}
\caption{Velocities of spin excitations 
as a function of the magnetic field.}
\label{velo}
\end{figure}

\begin{figure}
\caption{
Critical exponents  $\alpha_{\rm s}^{\rm R}$ 
and $\alpha_{\rm s}^{\rm C}$ for oscillating sectors
in the spin correlator as a function of 
the  magnetic field.}
\label{exp}
\end{figure}

\begin{thebibliography}{99}
\bibitem{numeric} The cut-off parameters $\Lambda^{\pm}_\alpha$
 are determined so as to minimize the free energy.
\end{thebibliography}
\end{document}